\documentclass[]{aa}
\usepackage{graphicx}
\usepackage{graphics}
\usepackage{color}
\usepackage[varg]{txfonts}
\usepackage{epstopdf}
\usepackage{multirow}

\usepackage{natbib}
\begin{document}

\title{Comparative analysis of solar radio bursts before and during CME propagation}

\author{G.Dididze\inst{1,2}, B.M. Shergelashvili\inst{2,3,4}, V.N. Melnik\inst{5}, V.V. Dorovskyy\inst{5}, A.I.
Brazhenko\inst{6},  S. Poedts\inst{1}, T.V. Zaqarashvili\inst{2,3,7}, M.Khodachenko\inst{3,8,9}
}
\institute {Centre for mathematical Plasma Astrophysics, KU Leuven, Celestijnenlaan 200 B, B-3001, Leuven, Belgium\\
             \and
             Abastumani Astrophysical Observatory at Ilia State University, Cholokashvili Ave. 3/5, Tbilisi, Georgia\\
                \and
             Space Research Institute, Austrian Academy of Sciences, Schmiedlstrasse 6, 8042 Graz, Austria\\
             \and
             Combinatorial Optimization and Decision Support, KU Leuven campus Kortrijk, E. Sabbelaan 53, 8500 Kortrijk, Belgium\\
             \and
             Institute of Radio Astronomy, National Academy of Sciences of Ukraine, Kharkov, Ukraine \\
             \and
             Poltava Gravimetric Observatory within S.I. Subbotin Institute of Geophysics, Poltava, Ukraine\\
             \and
             Institute of Physics, IGAM, University of Graz, Universit\"atsplatz 5, 8010 Graz, Austria\\
             \and
             Skobeltsyn Institute of Nuclear Physics, Moscow State University, Moscow, Russia\\
             \and
             Institute of Astronomy of the Russian Academy of Sciences, 119017, Moscow Russia\\
}

\abstract
{As is well known, CME propagation often results in the fragmentation of the solar atmosphere on smaller regions of density (magnetic field) enhancement (depletion). It is expected that this type of fragmentation may have radio signatures.}{The general aim of the present paper is to perform a comparative analysis of type III solar and narrow-band type-III-like radio burst properties before and during  CME events, respectively. The main goal is to analyze radio observational signatures of the dynamical processes in solar corona. In particular, we aim to perform a comparison of local plasma parameters without and with CME propagation, based on the analysis of decameter radio emission data.}{In order to examine this intuitive expectation, we performed a comparison of usual type III bursts before the CME with narrow-band type-III-like bursts, which are observationally detectable on top of the background type IV radio bursts associated with  CME propagation. We focused on the analysis of in total 429 type III and 129 narrow-band type-III-like bursts. We studied their main characteristic parameters such as frequency drift rate, duration, and instantaneous frequency bandwidth using standard statistical methods. Furthermore, we inferred local plasma parameters (e.g.,\ density scale height, emission source radial sizes) using known definitions of frequency drift, duration, and instantaneous frequency bandwidth.}{The analysis reveals that the physical parameters of coronal plasma before CMEs considerably differ from those during the propagation of CMEs (the observational periods 2 and 4 with type IV radio bursts associated with CMEs). Local density radial profiles and the characteristic spatial scales of radio emission sources vary with radial distance more drastically during the CME propagation compared to the cases of quasistatic solar atmosphere without CME(s) (observational periods 1 and 3).}{The results of the work enable us to distinguish different regimes of plasma state in the solar corona. Our results create a solid perspective from which to develop novel tools for coronal plasma studies using radio dynamic spectra.}


\titlerunning{Comparative analysis of solar radio bursts}

\authorrunning{Dididze et al.}

\maketitle
\section{Introduction}
Solar radio emission is connected with solar magnetic activity, which maintains plasma acceleration and heating processes manifesting themselves in various dynamic events \citep{Mclean1985,Aschwanden2005}. The radio observational instruments are able to detect the related electromagnetic radiation in the form of different kinds of type II, type III, type IV radio bursts, and others \citep{MC1985,Mel2005}, carrying information about their source dynamic events. Among them, type III bursts are produced by energetic electrons, which propagate along magnetic field lines in the solar corona \citep{suzuki85,sinclair2014,Melnikkon08}. These particle beam sources of  type III bursts are thought to originate from eruptive processes in the solar atmosphere related to magnetic reconnection. On the other hand, type IV bursts are clearly associated with the onset and propagation of Coronal Mass Ejections (CMEs) \citep{stewart85,melnik08}. The types of radio bursts under consideration represent important diagnostic tools for the remote study of density profiles, properties of solar radio emission sources, and other physical characteristics of the solar atmosphere \citep{bougeret,Mann99,sinclair2014,melnikboiko11}.

Before we turn to the analysis of the observations, it is important to note that type III radio bursts are believed to be excited by several known plasma generation mechanisms \citep{Ginzburg58}. It is supposed that fast electrons generate Langmuir waves along their propagation. Furthermore, the excited Langmuir waves ($l$) can be involved in dynamic processes that result in the excitation of electromagnetic waves, with a characteristic frequency close to  that of local electron plasma. On the other hand, it is well known that meter and decimeter bands of radio emission spectra of  type IV bursts are characterized by a clear internal structure manifested through the so-called zebra pattern of spectrum and fiber bursts \citep{chernov2006,stewart85}.
It is conventionally accepted that one of the possible mechanisms for the creation of fiber bursts involves three wave-wave interaction processes of the form $l+w \rightarrow t$, where $w$ denotes the low frequency whistler modes \citep{rausche,chernov,kuijpers}.
The physical mechanism of the excitation of type IV radio bursts is related in most cases to the onset of CMEs.

The main focus of this paper is the observational study not of the fiber burst or zebra pattern structures of  type IV bursts, but instead the modified (narrow banded) type-III-like bursts within the decameter range of the spectrum that covers
the characteristic frequencies $8-32\;$MHz. These modified type-III-like bursts are detectable on top of type IV busts and seemingly undergo drastic influence from the latter. It is also well known that this emission most plausibly originates from higher altitudes in the solar atmosphere with rather lower density plasmas compared to its values at the solar surface. In agreement with previous studies \citep{gopals,antonov14}, we assume that the structured dynamic spectrum develops on top of background type IV radio bursts in the decameter range and can be a representation of radio bursts with similar morphological properties to usual type III bursts. Their spectral properties are significantly modified by the CME passage through the atmosphere and examination of this expectation is the main topic of current investigation. These specific radio bursts can be named as modified or narrow-band type-III-like bursts. Hereinafter, we refer to them as narrow-band type-III-like bursts. In the literature a number of other non-standard type III radios bursts such as type IIIb or type IIIl have been identified\citep{sinclair2014}. There was no significant number of type IIIb in our observational period of time, therefore we omitted them from the analysis. As regards the type IIIl bursts, according to \citet{cane2002}, are observed in low frequencies usually below $10\;$ MHz. Their duration can be several tens of minutes, which is significantly longer than the durations of our usual type III bursts. These bursts \citep{cliver2009} are associated to type II bursts, which were absent in our observational period. Therefore, the modified nosebanded type III bursts observed by us within the range $14-32\;$MHz cannot be identified as type IIIl bursts. In our case, the dynamic picture we have in mind is based on the assumption that there are regular sources of accelerated electron beams, which are generated mostly in the reconnection processes of closed magnetic field structures at lower altitudes (significantly far from our radio observational window). These electron beams then travel through the ambient plasma, in one case without a CME and in the other case with the presence of a CME, then they reach the heights in which we are interested (8-32 MHz observational window). This way in either of the cases the electron beams play the role of plasma probing agents, which produce a sequence of the type-III-like radio bursts with different spectral properties before, during, and after the CME. Consequently, the key object of the current study is to perform a systematic comparison of the properties of the emitted type III bursts in those consecutive intervals of time. Hence, we base our further conclusions bearing in mind the fact that the type III bursts observed in the initial quasi-stationary atmosphere and during the CME propagation (with background type IV radio burst) could be characterized with similar parameters like frequency drift, duration, and instantaneous frequency bandwidth \citep{chernov2007,mlkonov08,koval2010}, and these parameters contain the information about the coronal and interplanetary plasma in which the CME is propagating.

The radio telescope Ukrainian Radio interferometer of National Academy of sciences--2 (URAN--2) observed a type III burst storm and type IV bursts on June 13, 2014. In the present paper, a comparative analysis of the mentioned different states of the solar atmosphere is carried out with the aim to detect and measure properties of the coronal plasma before, during, and after the CMEs. As a matter of fact, solar radio observations are used to explore the solar wind properties needed for the modeling of solar density profiles \citep[e.g., see][]{Mann99,newkirk61} and heating and/or acceleration processes \citep[e.g.,\ see][and references therein]{shergelashvili}. Another important aspect of the analysis carried out in this paper is the combination of the radio data with the optical observations. The research results reported here are an attempt to analyze solar radio emission, exploiting the radio dynamic spectra (patterns from the ground-based telescope URAN-2 (Ukraine, Poltava), see Fig.~\ref{spectri}), in combination with the CMEs' position and time evolution inferred from data from the space-based optical telescope The Large Angle Spectroscopic Coronagraph (LASCO) onboard of Solar and Heliospheric Observatory (SOHO); see Fig.~\ref{CME}). We observed a statistically significant number of radio bursts and investigated their parameters. Our observations and statistical analysis enabled us to determine the plasma density profiles and radial sizes of the emitting sources, before the onset and along the CME paths.
\section{Observations and data analysis}\label{observations}

\begin{figure*}[h]
\begin{minipage}[h]{0.7\linewidth}
\center{\includegraphics[scale=1.4]{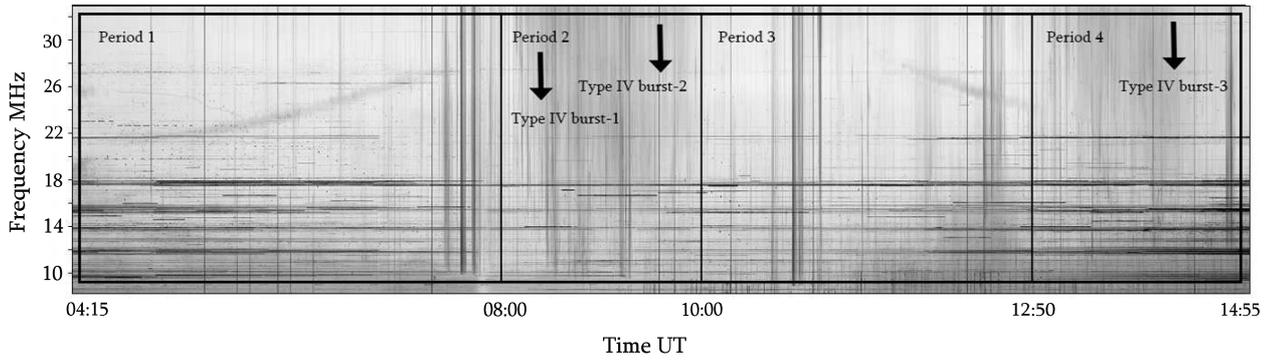}}
\end{minipage}
\caption{Observational data (dynamic radio spectrum) of June 13, 2014  from 04:15 UT to 14:55 UT by URAN-2 (Poltava, Ukraine). Period~1 corresponds to the time interval 04:15 - 08:00 UT, period 2 to 8:00 UT - 10:10 UT, period 3 to 10:10 - 12:50 UT, and period 4 to 12:50 UT - 14:55 UT.}
 \label{spectri}
 \end{figure*}

\begin{figure*}[h]
\begin{minipage}[h]{0.65\linewidth}
\center{\includegraphics[scale=0.24]{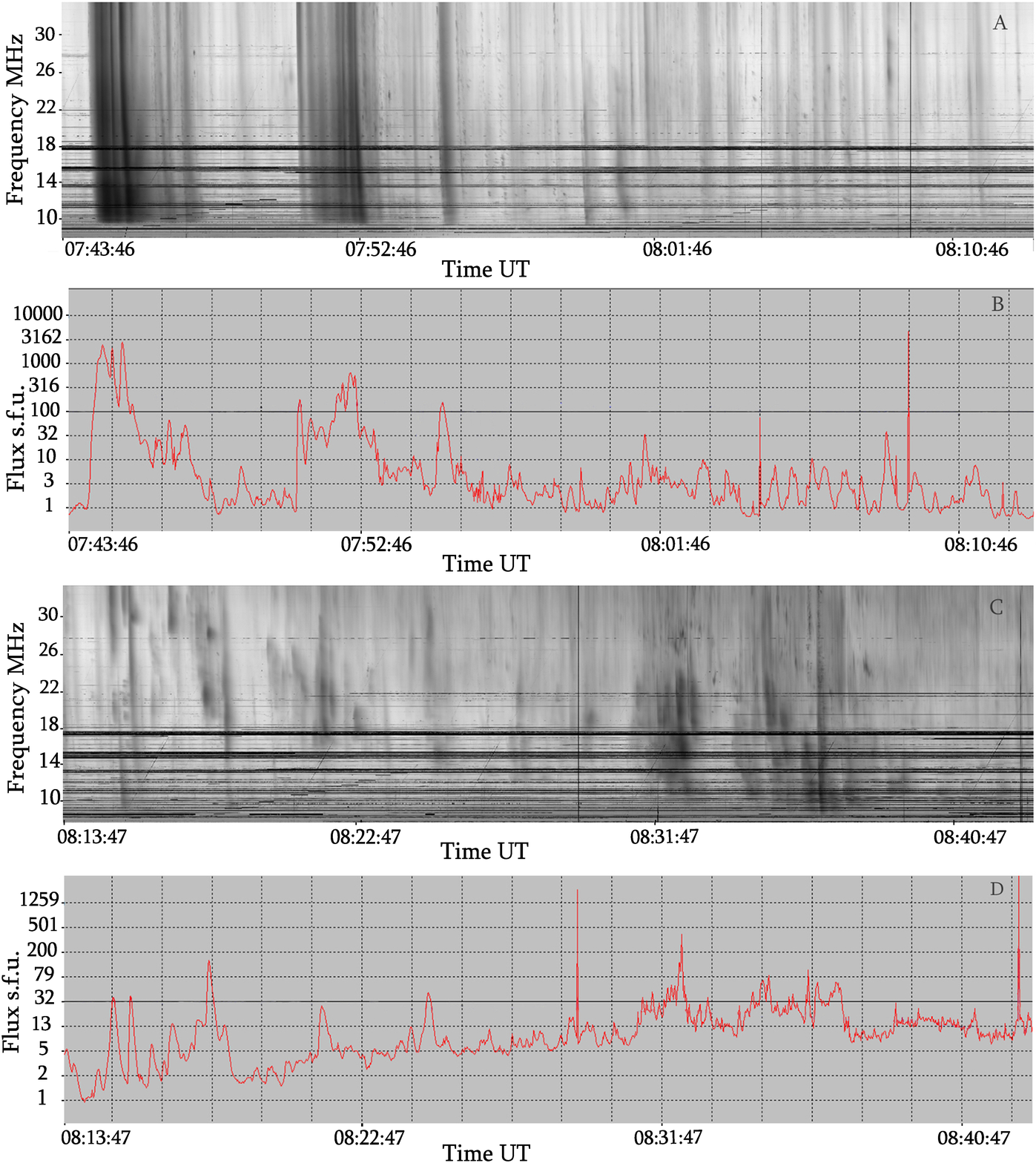}}
\end{minipage}
\caption{ Examples of the decameter type III and narrow-band type-III-like bursts. The dynamic radio spectrum of the type III burst (panel A) and variation of its flux (panel B) in time. The dynamic radio spectrum of the narrow-band type-III-like bursts (panel C) and variation of their fluxes (panel D) in time.}
 \label{burstsample}
 \end{figure*}

The type III and narrow-band type-III-like radio bursts discussed in this paper have been analyzed using data from the radio telescope URAN-2 (Poltava, Ukraine). This instrument represents the phased array of effective area $28000\;$m$^2$ consisting of $512$ broadband cross dipoles operating at frequencies of $8-32\;$MHz. The observation time span covered from 04:15 UT until 14:55  UT of June 13, 2014. We selected this date because it was very rich in dynamical events and directly corresponded to our objectives, that is,\ we had periods without type IV bursts but populated with type III radio bursts, and periods including type IV radio bursts representing background for (as we call them) narrow-band type-III-like bursts, which have similar spectral morphological properties as usual type III bursts, but with significantly different characteristic values of their dynamic parameters, and are observed in periods with CME onset. In total we analyzed $558$ radio bursts, $429$ of which were usual type III bursts and the other $129$ are modified by CME propagation.
We divided the total time span of the observation into four consecutive periods. We selected two time intervals with no CME propagation: the window from 04:15 to 08:00  UT we call period~1, and, analogously, we have period~3, which lasted from 10:10 until 12:50 UT. Period~3 represents the time span following the two coupled type IV bursts of period~2, and one before the third type IV burst occurred in period~4.
Accordingly, we distinguish two time intervals where type IV-like patterns are observed:\ from 8:00 UT  until 10:10 UT (period~2) and from 12:50 UT until 14:55 UT (period~4). Furthermore, throughout the text the type III bursts observed in periods 1 and 3 are referred to as usual type III bursts and those observed in periods 2 and 4 as narrow-band type-III-like bursts. The difference in corresponding dynamical spectra is discussed below in connection with Fig.~\ref{burstsample}. Consequently, we were able to investigate the radio emission properties of bursts in observational periods preceding, during, and following the CMEs that occurred within periods 2 and 4 (see the picture of the whole observational day in Fig.~\ref{spectri}) and to perform a comparison of the statistics of the radio events within the different time intervals.

\begin{figure*}[h]
\begin{minipage}[h]{0.7\linewidth}
\center{\includegraphics[scale=0.2]{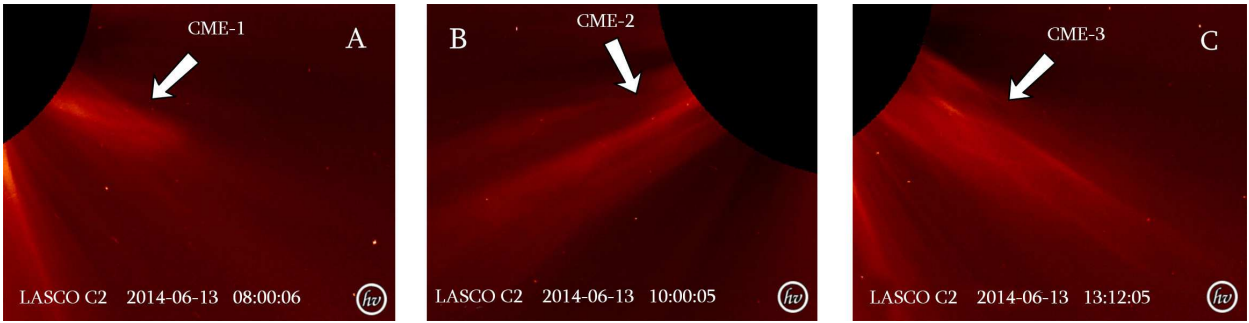}}
\end{minipage}
\caption{CMEs detected by the Solar and Heliospheric Observatory (SOHO) during our observation periods (source: www.helioviewer.org), which caused type IV bursts. Panels A (CME 1) and B (CME 2) correspond to period 2, and panel C (CME 3) corresponds to period 4.}
 \label{CME}
 \end{figure*}

At the same time, a number of space-based optical telescopes such as the SOHO and the Solar Dynamics Observatory (SDO)  detected  two CMEs, namely CME~1 with start time 07:36 UT, (see panel~A in Fig.~\ref{CME}, source \url{www.helioviwer.org}), and  CME~2 with start time 08:24 UT (see panel~B in Fig.~\ref{CME}), during period~2. In period~4 another CME was detected, which we call CME~3, with start time  12:24 UT (see panel~C in Fig.~\ref{CME}).
Therefore, we associate these CME events with the type IV radio bursts we observed. Also, we interpret the radio spectra within period~2 as the double pattern structure of the type IV radio bursts. Hence, we subdivided period~2 into two sub-intervals:\ 8:00-9:00 UT, corresponding to CME~1, and 9:00-10:10 UT, corresponding to CME~2. This subdivision has been inspected by us during the analysis of frequency drift and durations and other parameters for narrow-band type-III-like bursts. As we show in panel A (blue dashed and red solid lines) of Fig.~\ref{driftgraph}, when we divide period 2 into two equal parts and draw the approximations of drift rate dependence on the frequency, we observe that characteristic values of drift rates differ from each other drastically. The drift rates within time interval 08:00 to 09:00 (blue observed datapoints and dashed line of linear regression) have about 1.5 times larger values than those in interval 09:00-10:10 (red observed datapoints and solid line of linear regression). We obtain similar significant differences of the magnitude and/or slope of the regression lines within these sub-periods for their durations and other physical quantities as well (see panel A in Figs.~\ref{durationgraph}, \ref{bandwidthgraph}, and \ref{rLbound}). These findings justify and give an indication that our subdivision of period 2 into two parts is reasonable, even though because of the overlap of the time spans of the CME1 and CME2, the type IV radio bursts associated with them are not immediately visually distinguishable in the dynamical spectrum. Our argumentation is further proved by comparison with period 4, where we have only a single CME. In the latter case, when we formally divide period 4 into two equal intervals such drastic differences in the properties of physical quantities do not appear (characteristic rates and regression line slopes are of the same order of magnitudes), what indicates that period 4 is occupied by only a single type IV radio burst. Further details of quantitative analysis can be found in corresponding sections below.
In general, we identify both period~2 and period~4  as periods of type IV radio bursts (CME onset and propagation), within which the narrow-band type-III-like bursts carry the information about the observed CMEs and state of the solar atmosphere.

In panel~A of Fig.~\ref{burstsample} we show examples of a type III radio burst observed within period 1 or 3. According to recent publications \citep{melnik08,melnikboiko11} it is conventionally accepted that the type III burst fluxes change over a very wide range,\  from a few solar flux units ($1\;$s.f.u.$\;=10^{-22}\;$W m$^{-2}$ Hz$^{-1}$) up to $10^6\;$s.f.u. In agreement with previously reported measurements, our observations also evidenced that in periods 1 and 3, corresponding to non-disturbed corona, the type III radio bursts are observed within wider frequency bands compared to the ones seen in periods 2 and 4. Besides, we estimated the corresponding total fluxes and observed that the average values of type III bursts vary between $6.3-6.7$ s.f.u. in respective periods 1 and 3, excluding rare instances of powerful type III bursts with fluxes of thousands of solar flux units. We plot the time profile of the flux corresponding to the burst shown in panel~B. On the other hand, narrow-band type-III-like bursts observed in periods 2 and 4 show mean characteristic larger fluxes of magnitude compared to periods 1 and 3, which is in good agreement with previously obtained data \citep{antonov14}. As we mentioned above, the narrow-band type-III-like bursts are seen on top of the background type IV bursts and represent the ground physical state of corona during CME propagation through it. In panel~C of the same figure, we demonstrate the part of the dynamic spectrum from period~2 that shows a set of narrow-band type-III-like bursts. As regards the corresponding total fluxes, their mean values are many times larger than ones for type III bursts in periods 1 and 3, and they amount to about $60-65$ s.f.u in period~2 and $30$ s.f.u in period~4, accordingly. Finally, we discovered narrow-band type-III-like bursts with positive drift rates as well, which could become the subject of a separate analysis complimentary to this current study. All these observed properties of the bursts are in good agreement with other similar measurements \citep{antonov14,mlkonov08}.

\begin{figure*}[h]
\begin{minipage}[h]{0.65\linewidth}
\center{\includegraphics[scale=0.8]{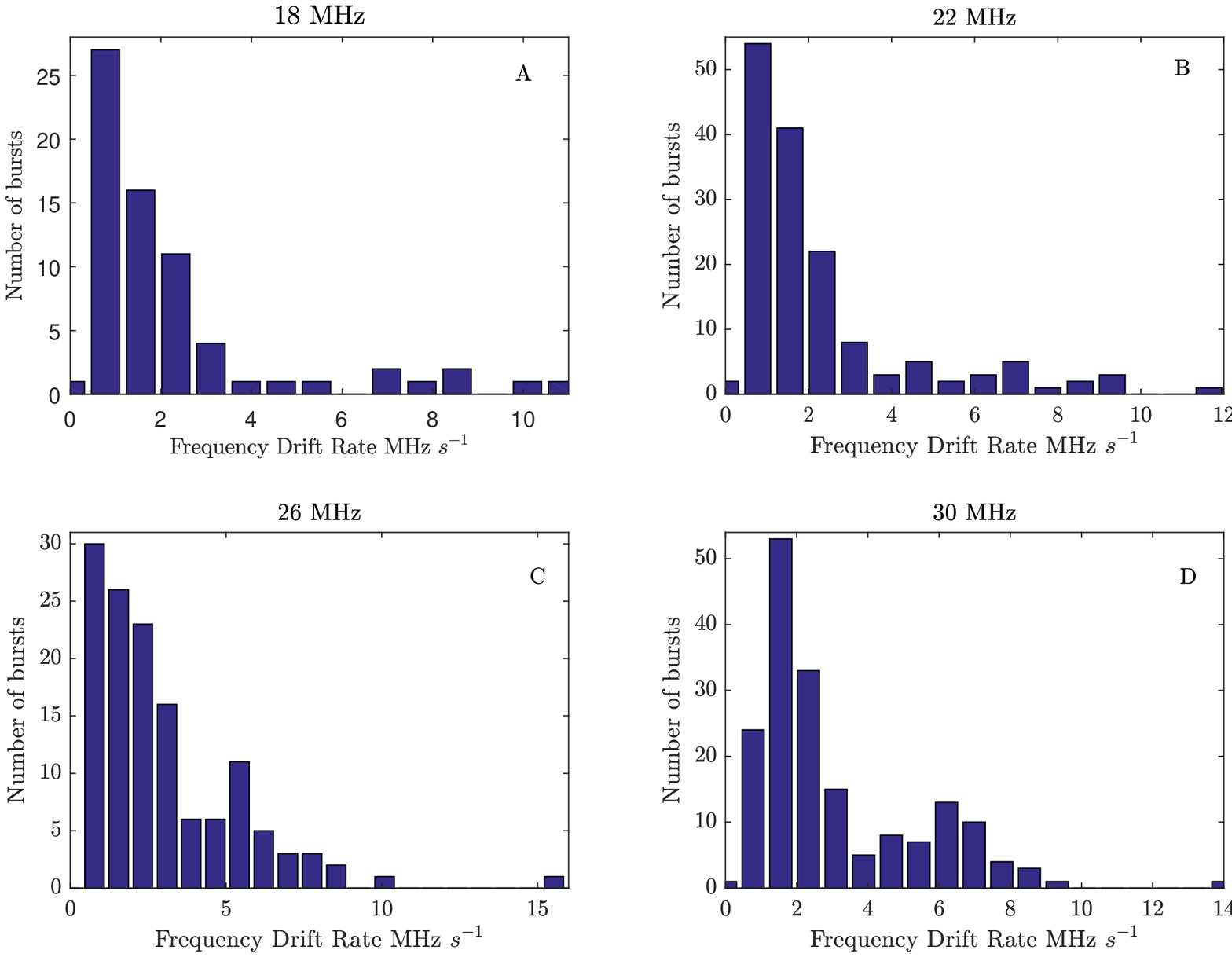}}
\end{minipage}
\caption {Example of drift rate histograms for the type III bursts during period 1 at frequencies $18\;$MHz (A), $22\;$MHz (B), $26\;$MHz (C), and $30\;$MHz (D).}
 \label{histdriftrate}
 \end{figure*}

In our analysis we follow the strategy that comprises direct measurements of the frequency drift and duration rates and uncertainty errors of observed radio bursts in each observational periods. Then the rates of the instantaneous frequency bandwidth (IFB) and characteristic length scales of the radio emission sources (with errors) are evaluated by using these measurements of the drift rates and/or durations (see corresponding descriptions in subsections below). As is known, within the framework of the accelerated particles beam mechanism, the emission frequency is related  to the local plasma frequency $f_{pe} =\left ( {e^2 n_e}/{\pi m_e}\right )^{{1}/{2}}$,
where $e$, $n_e$ , and $m_e$ are the electron charge, the electron number density, and its mass, respectively.
We divided the observational frequency domain $8-32\;$MHz into six sub-bands: $8-12\;$MHz, $12-16\;$MHz, $16-20\;$MHz, $20-24\;$MHz, $24-28\;$MHz, and $28-32\;$MHz. Before we analyze the results of measurements, it should be noted that we were not able to find
a statistically meaningful number of bursts within the range $8-12\;$MHz. Therefore, we detected the physical quantities for the frequency sub-bands  only starting from $12\;$MHz and beyond. Then we calculated the occurrence rates of the frequency drift, duration, and IFB values within each mentioned frequency sub-band of the dynamical spectrum and plotted corresponding histograms  For the sake of brevity we demonstrate only one example of such a set of histograms for the case of the frequency drift for period 1 in Fig.~\ref{histdriftrate}.  For the frequency drifts, durations, and IFBs, similar procedures are applied in each observational period separately and corresponding histograms constructed. In order to derive the mean  values for each frequency sub-band, we performed a Gaussian function fit to the plotted histograms, variance of which also determined the original observational uncertainty errors of the detected mean rates (further details about the total error estimation are given below).

In Figs.~\ref{driftgraph}-\ref{rLbound}, we plot observed values of quantities and fitting lines for periods 1 and 2 in corresponding panels A, and for periods 3 and 4 in panel B, in order to make states of the solar atmosphere before and during the CME directly comparable to each other. It should also be  noted that each observed value shown by the asterisks in Figs.~\ref{driftgraph}-\ref{bandwidthgraph} is calculated as a mean value of the Gaussian fitted to the corresponding histogram. Also the uncertainty errors are calculated. These errors come from the variance of the fitted Gaussian and the linear regression performed over the computed mean values and are shown in Figs.~\ref{driftgraph}-\ref{bandwidthgraph} as black and colored lines, respectively. We take into account the additional uncertainty coming from the linear regression (and width of the observational frequency sub-band in a dynamical spectrum) as we consider in cases of drift rate (Fig.~\ref{driftgraph}) and IFB (Figure~\ref{bandwidthgraph}) the linear regression, which is rather rough for the measured mean values of quantities but still enough for the analysis carried out in this paper. Therefore we compromise on taking linear fittings instead of higher order polynomial or other functions, but we impose the extra uncertainty on top of the original one that come from the histograms. A similar procedure is taken for durations (Fig.~\ref{durationgraph}), however in this case the linear fitting is applied to data and frequencies in logarithmic scales as the final function we are looking for is the power law.

More specifically, for the error estimation in the measurements of drift rates and IFB we use the  expression
\begin{equation}\label{errordriftifb}
\varDelta Y _{drift,IFB}= \overline{\varDelta {Y_{obs}} +p_1 \varDelta f} \frac{\left \lvert Y_{obs}-p_1 f-p_2\right \rvert }
{\overline{\left \lvert Y_{obs}-p_1 f-p_2\right \rvert}},
\end{equation}
where $p_1$ and $p_2$ are the coefficients (their values are given in corresponding tables below) of the linear regression and overlines on top of the expressions denote the simple mean values of them.
In case of duration, we use the expression
\begin{equation}\label{errorduration}
\varDelta Y_{Duration} = \varDelta {Y_{obs}} + q_1\varDelta f\left \lvert  \frac{Y_{obs}}{f} +\frac{\varDelta {Y_{obs}}}{f} \right \rvert \exp \left ( \varDelta {\ln Y_{obs}}\right ),
\end{equation}
where,
\begin{equation}\label{erroryln}
\varDelta {\ln Y_{obs}} = \frac{\left \lvert \ln Y_{obs}-q_1 \ln f-q_2 \right \rvert}{\overline{\left \lvert \ln Y_{obs}-q_1 \ln f-q_2\right \rvert}},
\end{equation}
where $q_1$ and $q_2$ are the coefficients of the linear regression of duration vs. frequency, both in logarithmic scale.
Finally, we evaluate the errors of the emission source sizes in Fig.~\ref{rLbound} as
\begin{equation}\label{errorysource}
\varDelta Y_{Source} = 0.3c\varDelta Y_{Duration}.
\end{equation}
In all these expressions, $\varDelta f=2\;$MHz represents the half widths of the above-mentioned frequency sub-bands in dynamic the spectrum. Equipped with the outcomes of the data analysis procedures outlined so far, we turn to the description of the results for each physical parameter observed in further detail.


\subsection{Drift rate}\label{driftrate}

The drift rate is one of the most important parameters of solar radio bursts. It corresponds to the time derivative of the emission frequency $df/dt$ (MHz s$^{-1}$) at a given frequency level. According to the definition, we calculated the frequency drift rates as a slope of local lines connecting the burst peak times at the boundaries of each frequency sub-band in the frequency-time diagram (local, piecewise linear fitting procedure).
Then, we performed a linear regression procedure on the sequence of the mean values of the frequency drift in different frequency sub-bands (as outlined above) obtained by the Gaussian fit on the histograms.

\begin{figure*}
\begin{center}
\includegraphics[scale=0.68]{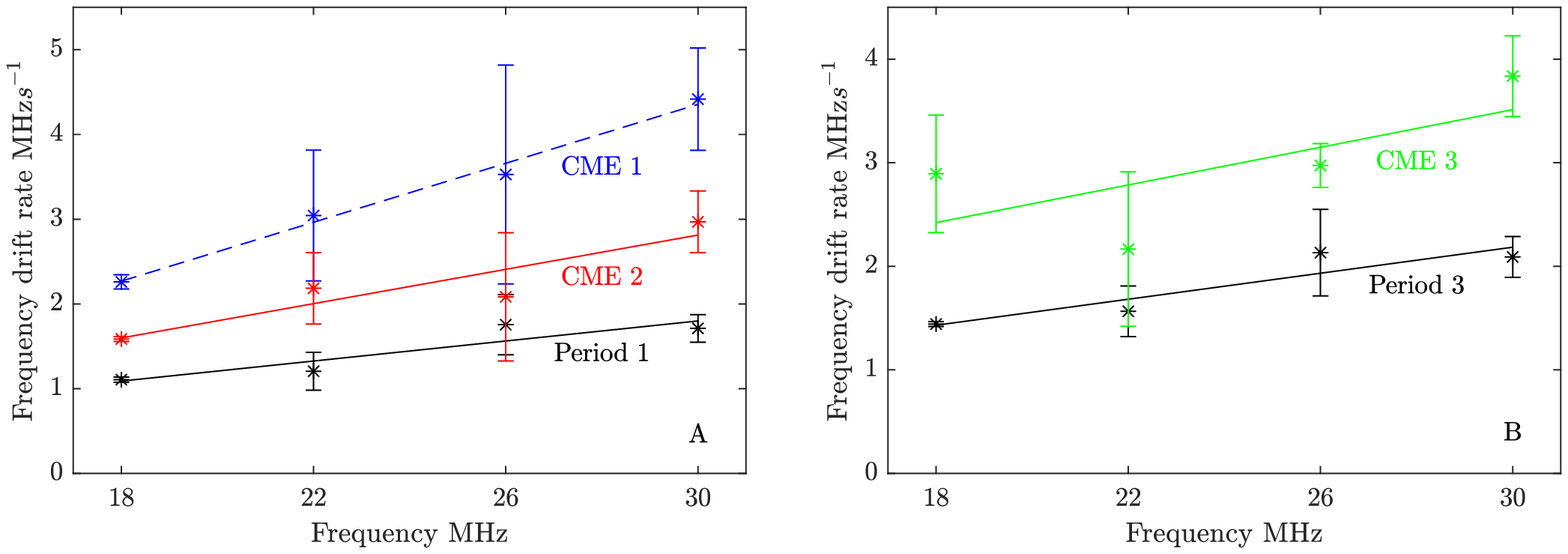}
\end{center}
\caption{Drift rate dependence on frequency according to the observations of type III and narrow-band type-III-like 
bursts on June 13, 2014. Panel A corresponds to periods 1 and 2,
 and panel B to periods 3 and 4.
In panel A the dashed line linear regression with corresponding observationally measured values (asterisks with error 
bars) are shown in blue and they are related to the interval 08:00-09:00 UT when  CME~1 was detected, and similarly the 
solid line and corresponding measured values are shown in red for the interval 09:00-10:10 UT, when CME~2 was detected. 
In panel B again the solid line and corresponding measured values are given in green and they show results for the time 
interval 12:50 UT-14:55 UT, when CME 3 was detected. In both panels for data  and lines in period 1 and period 3, black 
is used.}
\label{driftgraph}
\end{figure*}

We approximated the dependence of the drift rate on the emission frequency for observed
bursts by a linear fitting
\begin{equation}\label{linear}
 \frac{df}{dt} = -A \cdot f+B,
\end{equation}
 where coefficients $A$ (s$^{-1}$) and $B$ (MHz s$^{-1}$) are to be evaluated through linear fitting to
drift rates evaluated in each frequency sub-band, $f$ (MHz)
(in all following formulas throughout the text we measure the emission frequency $f$ in MHz)
and the corresponding fitting lines are plotted in Fig.~\ref{driftgraph}.
In panel~A of Fig.~\ref{driftgraph}, two lines are plotted corresponding to the sub-intervals within which type IV radio bursts occur. The dotted blue line corresponds to the sub-interval 08:00-09:00 UT, when CME~1 was detected, while the solid red line corresponds to CME~2 in the sub-interval 09:00-10:10 UT.
Our results show that the mean drift rates for the type III bursts we observed fall between $1-2.5\;$MHz s$^{-1}$ and for the narrow-band type-III-like bursts they range between $1.5 - 4.5\;$MHz s$^{-1}$. The drift rates increase with the frequency in each period. In some cases, the drift rates can be as small as $0.4\;$MHz s$^{-1}$, and sometimes single bursts have drift rates of more than $10\;$MHz s$^{-1}$ \citep[which is in agreement with the similar result reported by ][]{melnik08}. In addition, we detected a difference between the values of the drift rates of the narrow-band type-III-like bursts in period 2, when we observed two CMEs with different properties.

\begin{table*}[h]
\caption{Values of the constants A and B of the linear approximation for the drift rate as defined in Eq. \ref{linear}.}
\centering
\begin{tabular}{ l  c  c  c  c c }

 Coefficients               & Period 1   & Period 2 (CME1) & Period 2(CME2) & Period 3 & Period 4 \\
\hline
\hline
 A, s$^{-1}$                & 0.06 &   0.17             &  0.10 &       0.06 & 0.09  \\

 B, MHz$\cdot$s$^{-1}$      &  0.02&  -0.86              & -0.22 &      0.30  & 0.79 \\

\end{tabular}\label{tabledrift}
\end{table*}

In Table~\ref{tabledrift}, the linear approximation coefficients are shown for each considered period.
The values of coefficient $A$ in  periods 2 and 4 are significantly larger than those in  periods 1 and 3. The physical meaning of this fact will be described below.
\citet{wild1950} reported that the drift rates of type III bursts observed within the frequency range
of $70-130\;$MHz can be linearly approximated with characteristic values of the coefficient $A$ varying within
the range $0.16-0.5\;$s$^{-1}$ , which is in good agreement with recent estimations reported by \citet{melnikboiko11}. In a similar study \citep{antonov14} the frequency drift was approximated by the power law function $\sim af^n$. In that work the characteristic values obtained for the constants were $a=7 \cdot 10^{-3}$ and $n=1.73$, which correspond to the value of the coefficient in our linear approximation $A=0.14\;$s$^{-1}$  for the case of no CME, while in the presence of type IV bursts the value of $n=1.23$ and $a=4.5 \cdot 10^{-2}$ gives a corresponding value $A=0.12\;$s$^{-1}$. As one can see from Table~\ref{tabledrift}, in  periods 1 and 3 we infer values that are a factor of about 0.43 smaller, and in periods 2 and 4 the values obtained by us are comparable with \citet{antonov14} estimations.
The mean frequency drifts are monotonically increasing functions of the radio emission frequency, meaning decreasing with distance from the reference point where the frequency is about $30\;$MHz. In general, the direction of the emission source motion can be oblique to the density radial gradient, but as we consider a large number of radio bursts, we assume that the angle between these two directions is distributed somewhat homogenously and therefore we can represent the average radial component of the source velocity by the expression

\begin{equation}\label{v_s}
 v_s =\frac{2}{\pi}\int_0 ^{\frac{\pi}{2}} v_0 \cos \alpha d \alpha,
\end{equation}
where $\alpha$ is the angle between source velocity and radial direction and $v_s$ is the mean source velocity.
In Eq.~(\ref{linear}) the inequality $|B/Af|\ll1$ is valid, which is justified by our measurement given in Table \ref{tabledrift}.
In our analysis the value of electron beam velocity is always fixed to $0.3c$ and there are recent similar measurements for high frequency radio emissions $80-240\;$MHz related to type III radio bursts from the solar limb sources \citep{mccauley2018}. These authors use the radio imaging technique, which allows them to not only infer the density profiles but also to directly measure the emission source velocity.
Thus we have
\begin{equation}\label{density}
 \frac{df}{dt} =f\cdot \frac{1}{2n_e}\cdot \frac{dn_e}{dr}\cdot v_s,
\end{equation}
where $r$ is the radial coordinate.  Using the last expression we evaluated the density profiles provided that the average source velocity $v_s$ is taken to be $0.3c$ for accelerated electron beams \citep{melnikboiko11}.
In such an analysis of the density profiles, either the shape of the profile has to be fixed by imposing existing known profiles and evaluating the corresponding velocities of the emission sources, or on the contrary fixing the mean value of the source velocity and deriving the shape of the density profile from the observed rates of the frequency drift. In the current study we used the second method as we were interested in a comparison of the actually observed density radial distributions before and during the CME, independently from preassumptions laying under of any known standard solar atmospheric profiles.
Based on the comparison of Eqs.~(\ref{linear}) and (\ref{density}), the following relation is valid:
\begin{equation}\label{A}
  A=-\frac{1}{2n_e} \cdot \frac{dn_e}{dr}\cdot v_s.
\end{equation}
Supposing  that  the  velocity of the source is given and $A$ is calculated from observations, Eq. \ref{A} can be used to infer the exponential density profile
\begin{equation}\label{n}
   n=n_0\exp \left (-\frac{2A}{v_s}r \right),
\end{equation}
where $n_0$ is a normalization factor representing the density rate at the reference altitude, where emission frequency is equal to $30\;$MHz.

\subsection{Duration}\label{duration}
\begin{figure*}
\begin{center}
\includegraphics[scale=0.7]{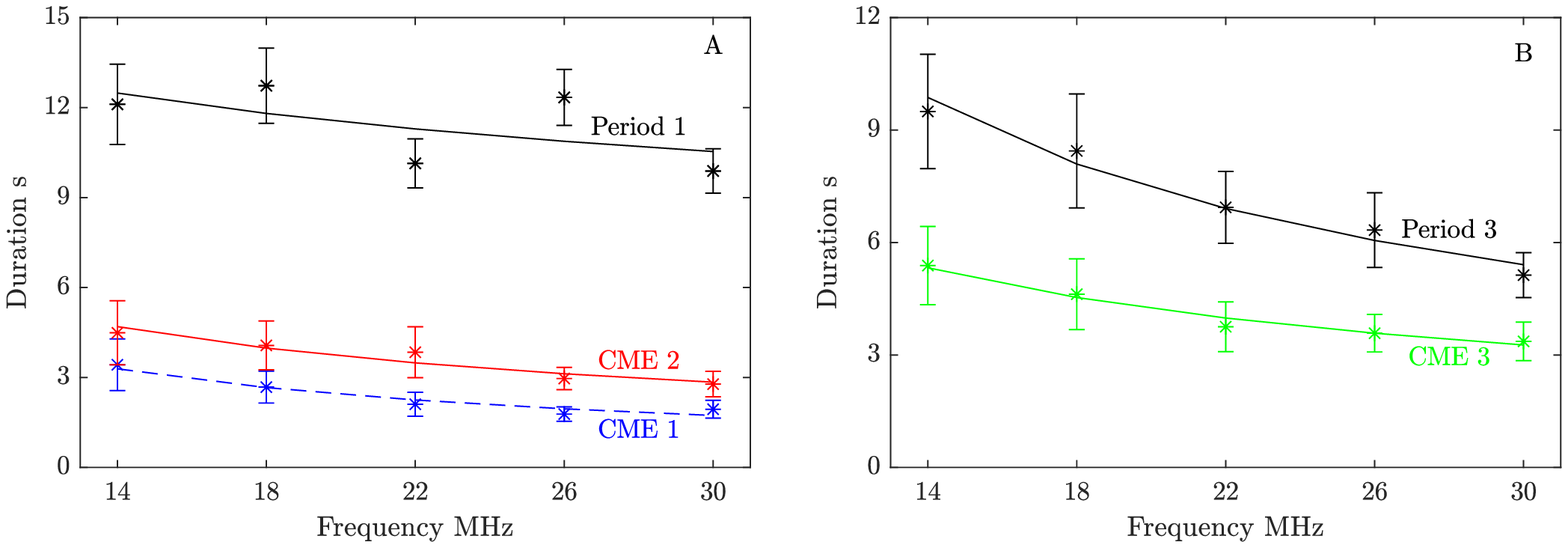}
\end{center}
\caption{Dependence of duration on frequency. The sequence of the line styles and coloring is the same as in Fig.~\ref{driftgraph}.}
\label{durationgraph}
\end{figure*}

The radio burst duration at a fixed frequency (i.e.,\ corresponding to a fixed density at a given radial distance) is caused by the radial sizes of the local radio emission sources and measured as half peak intensity levels \citep{suzuki85,elgaroy72,rutkevich}. Again we create data sets of the burst durations for each observational period separately, then we plot histograms of the duration occurrence rates and perform a Gaussian function fit to detect the mean duration values. There are several approximation formulas known for the dependence of duration on the emission frequency. In the frequency range 67$\;$kHz - 2.8$\;$MHz it has been approximated by \citet{suzuki85} as
\begin{equation}\label{suzukidulk}
    \tau= \frac{220}{f}=7.33 \left (\frac{f}{30}\right ) ^{-1},
\end{equation}
where the duration is measured in seconds.
On the other hand, \citet{elgaroy72} proposed  a dependence of the duration on the frequency by the  scaling law
\begin{equation}\label{lyngstad}
    \tau=60\cdot f^{-\frac{2}{3}}=6.21 \left (\frac{f}{30}\right ) ^{-\frac{2}{3}},
\end{equation}
\begin{table*}[h]
\caption {Values of the constants $k$ and $E$ of Eq. \ref{powerfunction}.}
\centering
\begin{tabular}{ c  c  c  c  c c }

 Coefficients   & Period 1   & Period 2 (CME1) & Period 2(CME2) & Period 3 & Period 4 \\
\hline
\hline
k              & -0.22    & -0.84         & -0.66         &  -0.79   & -0.64  \\

E, s                & 10.54     & 1.73             & 2.84            & 5.41      & 3.27 \\

\end{tabular}\label{durationtable}
\end{table*}
for the frequency range $300\;$kHz - $500\;$MHz. The durations obtained from our data are almost always smaller compared to Relation~\ref{suzukidulk}.
We approximate our measurements also as a scaling law similar to Expressions~\ref{suzukidulk} or~\ref{lyngstad},
 \begin{equation}\label{powerfunction}
 \tau= E \cdot \left (\frac{f}{30}\right ) ^{k},
 \end{equation}
where $\tau$ denotes the duration in seconds, $E$ is a constant measured in s, $f$ corresponds to the frequency, and $k$ is
the power exponent. We detect the values of the constant parameters from a fitting procedure outlined above and displayed in
Fig.~\ref{durationgraph}, and the obtained values of these parameters, $E$ and $k$, are given in Table~\ref{durationtable}.

\subsection{Instantaneous frequency bandwidth}\label{bandwidth_sec}
\begin{figure*}
\begin{center}
\includegraphics[scale=0.7]{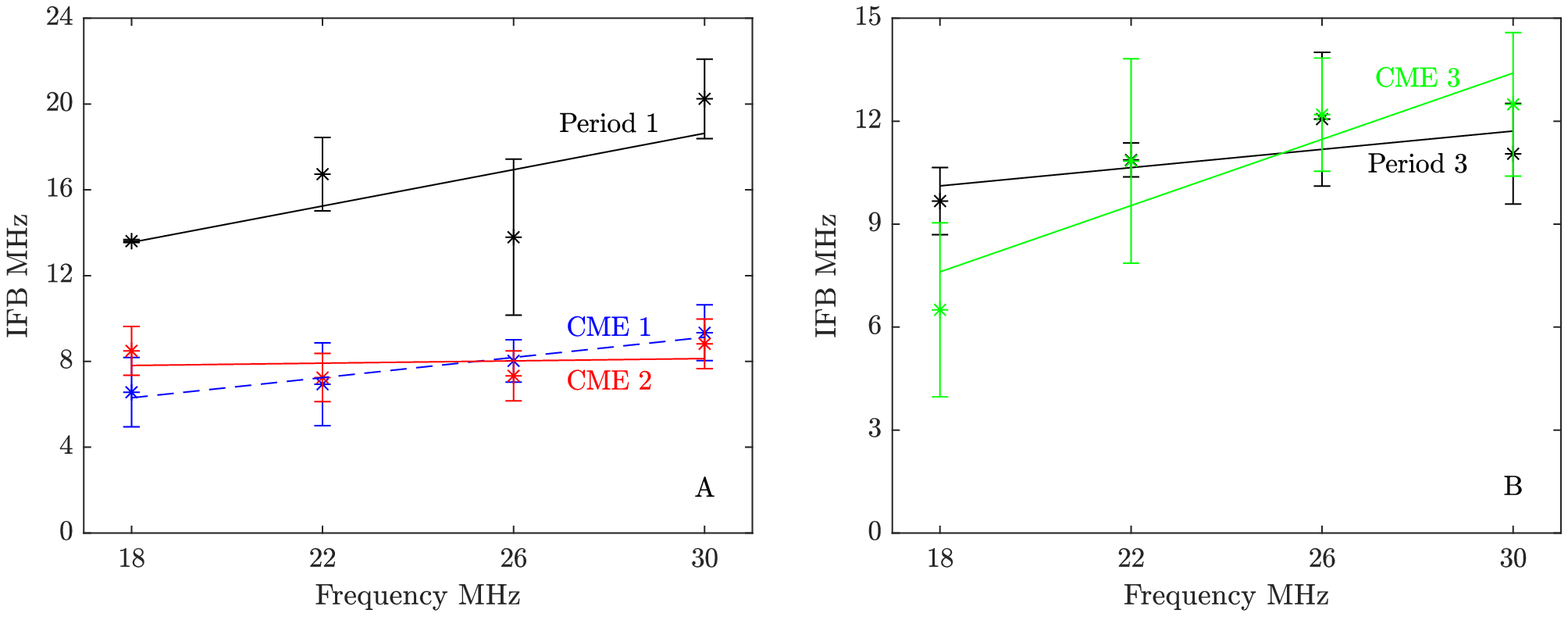}
\end{center}
\caption{Instantaneous frequency bandwidth dependence on the frequency. The sequence of the line styles and coloring is the same as in Fig.~\ref{driftgraph}.}
\label{bandwidthgraph}
\end{figure*}

Instantaneous frequency bandwidth is the additional measurable parameter that characterizes the properties of the radio bursts we investigate. It can be detected directly by inferring half power bandwidth, that is,\ the bandwidth is defined as the half width of the Gaussian fit of power in the frequency domain, at a given instance.
 There is another way to carry out IFB estimation by using the following relation with the burst duration and frequency drift:
 \begin{equation}\label{bandwidth}
    \Delta f =\frac{df}{dt} \cdot \tau.
 \end{equation}
In our observational data analysis we used the second method for the measurements: Expression \ref{bandwidth} for the evaluation of IFB.
This way we calculated the instantaneous frequency bandwidth values for each frequency sub-band and performed the same
statistical analysis as we did for drift rate and duration. Again, we first plotted histograms of the instantaneous frequency bandwidth distributions to derive the mean bandwidth values for each frequency sub-band. Next, we approximated the systematic dependence of the bandwidth on the emission frequency by a linear fitting,
  \begin{equation}\label{bandwidthapprox}
    \Delta f = C \cdot f + D,
  \end{equation}
where $C$ and $D$(MHz) are constants evaluated through fitting. The results of these approximations are shown in Fig. \ref{bandwidthgraph}.

\begin{table*}[h]
\caption {Regression coefficients  $C$ and $D$ of Eq. \ref{bandwidthapprox}.}
\centering
\begin{tabular}{ c  c  c  c  c c  }

Coefficients   & Period 1   & Period 2 (CME1) & Period 2(CME2) & Period 3 & Period 4 \\
\hline
\hline
 C             & 0.42     & 0.24          & 0.03        & 0.13   & 0.48  \\

 D, MHz         & 5.92     & 2.07          & 7.33         & 7.72    & -1.08 \\

\end{tabular}\label{bandwidthtable}
\end{table*}

The values of the fitting coefficients $C$ and $D$ are given in Table \ref{bandwidthtable} for each period. The sequence of panels, line, and marker styles and coloring in Fig.~\ref{bandwidthgraph} coincide with those in the figures for the drift rate and duration.
According to our observations, the IFB values for type III bursts lay between $9-21\;$MHz, and for narrow-band type-III-like ones they range from $6$ to $13\;$MHz. We find that the IFB is also an increasing function of the frequency, although for the case of CME2 it remains almost constant.

\section{Discussion}\label{discussion}
Our results show that the values of coefficient $A$ shown in Table \ref{tabledrift} range between $0.06-0.17\;$s$^{-1}$ , which is close to the results reported by \citet{wild1950} for the frequency range $70-130\;$MHz and analogously for the range $8-32\;$MHz reported by \citet{melnikboiko11}, where the values are in the range $0.07-0.12\;$s$^{-1}$. Within period 2 (as in period 4 the situation is different, as we argued above) we calculated two characteristic values of the coefficient $A$ and these values are drastically different from each other. We also observe that the values of coefficient $A$ in periods 1 and 3 are two to three times smaller than those in periods 2 and 4. We suppose that this is related to the physical properties of the two different CMEs that occurred in that period.  There may be different explanations of the observed properties of the local plasma. However, it is plausible to assume that significant sharpening of the density gradient in the instances of CME propagation may be related to certain magnetic islands, which represent local density enhancements due to confinement by the magnetic field that produces sharper density gradients at the edges between the islands and ambient coronal plasma. This hypothesis is in good agreement with the modern observations of magnetic islands and related density depletion regions \citep[for instance, see][]{erikson2014}.

It is not the primary goal of the current paper to build any rigorous density model as within our observational
framework we can estimate only the coefficient standing at $r$ in the exponential function of Expression \ref{n} or
equivalently the slope of the variation of frequency drift with the emission frequency determined by Relation \ref{A}.
Although, it could be appropriate to make a rough comparison of the values of the slope observed by us with ones recovered
from some of the existing coronal density models. For example, we analyzed two cases of such models given by
\citet{newkirk61}:
\begin{equation}\label{nnewkirk}
   n=n_{*}\cdot 10^{4.32\frac{R_{\sun}}{r}},
\end{equation}
where $n_{*}=4.2 \cdot 10^4\;$cm$^{-3}$  , and by \citet{leblanc1998}
\begin{equation}\label{nleblanc}
 n=\frac{n_{1}}{\left( r/R_{\sun}\right)^2} +\frac{n_{2}}{\left( r/R_{\sun}\right)^4} +\frac{n_{3}}{\left( r/R_{\sun}\right)^6},
\end{equation}
where  $n_{1}=3.3 \cdot 10^5 $, $n_{2}=4.1 \cdot 10^6$ , $n_{3}=8 \cdot 10^7\;$cm$^{-3}$.
We made a rough estimation of the corresponding values of exponent $A$ and compared them to the mean
observed ones. For this we recall that the $30\:$MHz and $20\:$MHz radio emission is approximately radiated from
altitudes $1.78R_{\sun}$ and $2.1R_{\sun}$, respectively. Therefore, from Newkirk's and Leblanc's models one
respectively has $A_{Newkirk}=0.22\;$s$^{-1}$, $A_{Leblanc}=0.23\;$s$^{-1}$ ($30\:$MHz) and
$A_{Newkirk}=0.16\;$s$^{-1}$, $A_{Leblanc}=0.19\;$s$^{-1}$ ($20\:$MHz).

The results also show that the type III burst mean durations in the frequency range $12-32\;$MHz are between $5-13\;$s \citep[which is in good agreement with values reported by][]{Mel2005}, and those for the narrow-band type-III-like bursts range between $1.5-5.5\;$s.

It is evident that the duration decreases with frequency in each period. Besides, we detected a difference between the values of durations of narrow-band type-III-like bursts corresponding to the two different type IV bursts observed within period 2, when two CMEs were launched from different heliographic positions.  As a matter of fact, the durations detected in the first sub-interval (08:00-09:00 UT, corresponding to CME~1) of period~2 have smaller mean values ($1.9-3.4\;$s) in each frequency sub-band compared to those in sub-interval 09:00-10:10 UT (corresponding to CME~2) ranging between $2.8-4.5\;$s.

 By inverting the electron density profile in Eq.~\ref{n}, an expression derived from the drift rates, and using the definition of the electron plasma frequency,
we can calculate the position altitude $h$ of the source propagating with respect to the reference point corresponding to $f^{(0)} _{pe}=30\;$MHz as
\begin{equation}\label{bidzina}
    h=\frac{v_s}{A} \cdot ln\left [ \frac{f^{(0)}_{pe}}{f_{pe}}\right ].
\end{equation}
This height is situated at different radial positions corresponding to local density scale heights (radial gradients) ultimately determined by the mean source velocity $v_s$. The parameter $A$ is the derivative of the drift rate with respect to the emission frequency, as defined in Sect. \ref{driftrate} and the values given in Table~\ref{tabledrift}.

As is well known, the burst durations carry information about the spatial size of the emission sources \citep{rutkevich12,rutkevich}. We calculate them using the expression
\begin{equation}\label{bidzi}
    L\approx v_s \cdot \tau=v_s \cdot E \left (\frac{f}{30}\right ) ^{k}
,\end{equation}
where $L$ is the radial length of the source.
\begin{figure*}
\begin{center}
\includegraphics[scale=0.7]{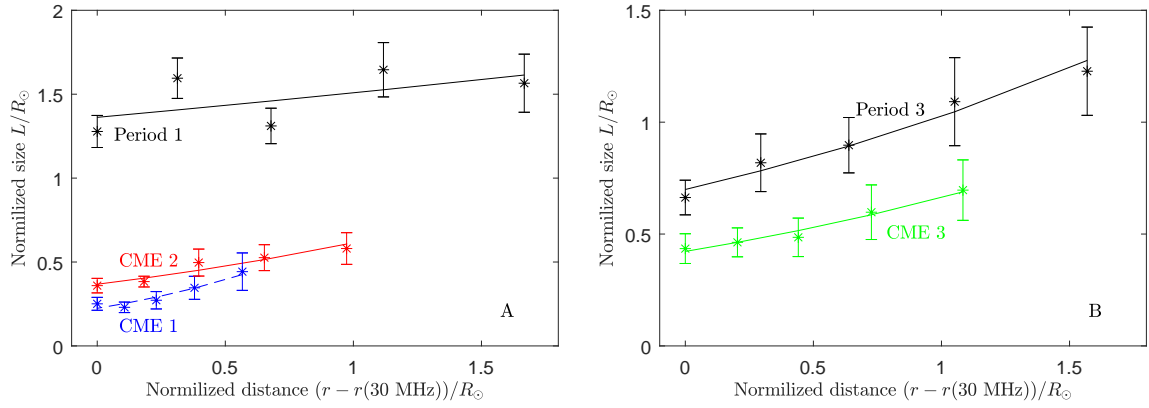}
\end{center}
\caption{ Dependence of the emission source sizes $L$ on the radial distance from reference point $r-r(30\;$MHz$)  $(both are measured in solar radii). The sequence of the line styles and coloring is the same as in Fig.~\ref{driftgraph}.}
\label{rLbound}
\end{figure*}

Figure~\ref{rLbound} displays the dependence of the radial length of the emission source on the radial distance (both normalized by $R_{\sun}$). The reference height corresponds to a frequency of $30\;$MHz and the emission source speed is fixed to $v_s=0.3c$ (particle beam excitation mechanism for both type III and narrow-band type-III-like bursts are taken into account, because of reasoning given in the introduction and further in the text).

\begin{table*}[h]
\caption {Density scale height $H_{n_{e}}$ and its fraction on average per observational period radio emission source size $\bar{L}$.}
\begin{tabular}{ l  c  c  c  c }

Quantity                        & Period 1       & Period 2(CME1 / CME2)& Period 3       &Period 4    \\
\hline
\hline
$\bar{L}=L/R_{\sun}$                & 1.48        & 0.31/0.47            & 0.94          &  0.54      \\

$H_{n_{e}}/ \bar{L}$, $R_{\sun}$    & 0.74       & 1.2/1.36              & 1.09           & 1.32      \\

$H_{n_{e}}$, $R_{\sun}$             & 1.09       &0.37/0.64               &  1.03         & 0.71        \\

\end{tabular}\label{sourcetable}

\end{table*}

In Fig.~\ref{rLbound} the length of the source increases with distance in each period. Also, the values of $L$ are significantly smaller in periods 2 and 4 (which range between $0.2-0.7\;R_{\sun}$) than in periods 1 and 3 (ranging between $0.65-1.65\;R_{\sun}$). As regards the values of radial widths of the observational window $h_w=r(14\;$MHz$)-r(30\;$MHz$)$, they are larger in the periods before the CME propagation (with sizes $h_w > 1.5\;R_{\sun}$ both for periods~1 and 3) compared to periods 2 and 4 with values $h_w\approx 0.6\;R_{\sun}$ (CME1), $h_w\lesssim 1\;R_{\sun}$ (CME2), and $h_w\gtrsim 1\;R_{\sun}$ (CME3). This means that the emission sources expand and reach their observed sizes at different radial distances depending on density profile differences between quiet (periods 1 and 3) and type IV radio burst (periods 2 and 4) epochs.
In Table \ref{sourcetable} we provide the values of the density scale heights $H_{n_{e}}$ and their ratios on average per period radio emission source size $\bar{L}$. In the periods of absence of CME (periods 1 and 3), the density scale height ratio on the mean emission source size $\bar{L}$ is rather small compared to the cases of CME propagation (periods 2 and 4). This fact may indicate that before the CME the characteristic sizes of the radio emission sources are large and they are comparable or even exceed the density scale height values (see Table \ref{sourcetable}). From the same table it is also evident that, on the contrary, during the CME's the emission, sources are considerably smaller than the characteristic length of the density radial variation. The smaller spatial scales of the emission sources and corresponding scale heights (Table \ref{sourcetable}) within periods 2 and 4 may indicate a link with the magnetic islands \citep{erikson2014} forming in the CME passage regions. On the other hand, narrowing of the radial observation windows demonstrates the sharper density radial gradients. The main final conclusion of this analysis is that both the density scale heights and emissions sources decrease during the CME propagation, however emission source sizes drop down more drastically, so that we observe the increase in the ratio of these quantities.

\section{Conclusions}\label{secconcl}
Our goal was to compare the characteristics of solar radio type III bursts during pre-CME eruption and CME propagation periods, as we wanted to observe whether there are any differences between these periods that are manifested in values of the plasma parameters. To accomplish this, we used data from the telescope URAN-2 recorded on  June 13, 2014. The results of our analysis show that, indeed, the plasma properties are significantly different in observation periods with and without CMEs.

The general tendency in the drift rate values we obtained is that they increase with the emission frequency, ranging between $1-2.5\;$MHz s$^{-1}$ for type III and  between $1.5-4.5\;$MHz s$^{-1}$ for narrow-band type-III-like bursts. The values of the linear fitting coefficient $A$ in periods 1 and 3 are two to three times smaller than those in periods 2 and 4. We concluded that this fact could be caused by the presence of magnetic islands \citep{erikson2014} at the CME passage areas.

The durations for type III bursts are between $5-13$ s and between $1.5-5.5 $ s for narrow-band type-III-like ones. The duration rate decreases with emission frequency in each period. Using the duration, we determined normalized spatial sizes of the emitting sources $L/R_{\sun}$ as a function of the normalized radial distance $h=r-r(30\;$MHz$)$. It is important to note that the positions of the reference points have been calculated using the density profile model for each period separately, corresponding to the respective values of the coefficient $A$. In pre-eruption periods, the characteristic sizes of the emitting sources range as $L=0.7-1.7\;R_{\sun}$. During the CME propagation periods, they range as  $L=0.2-0.7\;R_{\sun}$. In both cases, the emission sources expand with radial distance.

The instantaneous frequency bandwidth increases linearly with the emission frequency. Its values for type III bursts range between $9-20\;$MHz and for narrow-band type-III-like ones they lay between $6-12\;$MHz. The presence of the spread of the radio bursts in the frequency domain is usually ascribed to the range of the emitted radio wavelengths, which in turn is described in terms of the characteristic cross-sectional sizes of the magnetic structures.

The method presented in this work is useful for the radio diagnostics of pre-, during, and post- CME corona. We performed a general analysis of our results in conjunction with optical observational signatures of the CMEs. The results obtained in this paper create a solid background for the further development of observational and theoretical tools for coronal plasma diagnostics in future applications.

\begin{acknowledgements}
This work was supported by Shota Rustaveli National Science Foundation grants DI-2016-52 and FR17\_609. The investigation was initiated and developed within European Commission FP7-PEOPLE-2010-IRSES-269299 project - SOLSPANET. The work of G.D. has been realized within the framework of Shota Rustaveli National Science Foundation PHD student's research (individual) grant DO/138/6-310/14 and grant for young scientists for scientific research internship abroad IG/47/1/16. The work of B.M.S. and M.L.K. was supported by the Austrian Fonds zur Frderung der wissenschaftlichen Forschung (FWF) under projects P25640-N27, S11606-N16, and Leverhulme Trust grant IN-2014-016. The work of T.V.Z. was supported by the Austrian Science Fund (FWF)  under project P30695-N27. M.L.K. additionally acknowledges the support of the Russian Science Foundation (grant  18-12-00080) and FWF project I2939-N27. We would like to express our appreciation to Dr. C. Briand and Dr. J. Magdalenic for valuable discussions and constructive suggestions on the content of the manuscript. We are thankful to the referee, Prof. G.P. Chernov, for constructive comments on our paper that led to significant improvements of the content.
\end{acknowledgements}

\bibliography{mybibg}
\end{document}